\documentclass[12pt]{article}
\usepackage{verbatim}
\usepackage{amsfonts}
\usepackage{graphics}
\usepackage{amsmath}
\usepackage{times}
\usepackage{appendix}
\usepackage{color}
\usepackage{enumerate}
\usepackage{fancyhdr,latexsym,amsmath,amsfonts,amssymb,amsbsy,amsthm,url}
\usepackage[margin=0.5in,footskip=0.25in]{geometry}
\usepackage{graphics,graphicx,epsfig}
\usepackage{breqn}
\usepackage{caption,subcaption,float,subfloat}

\usepackage[english]{babel}
\usepackage[dvipsnames]{xcolor}
\usepackage{tikz}

\numberwithin{equation}{section}

\fancypagestyle{usual}{
  \lhead[\thepage]{}
  \chead[{\it \shortauthors}]{\sc \shorttitle}
  \rhead[]{\thepage}
  \lfoot[]{}
  \cfoot[]{}
  \rfoot[]{}
  
}

\makeatletter

\renewcommand{\section}{
  \@startsection
  {section}
  {1}
  {0pt}
  {1.1\baselineskip}
  {0.2\baselineskip}
  {\sc \centering}
}

\renewcommand{\subsection}{
  \@startsection
  {subsection}
  {1}
  {0pt}
  {1.1\baselineskip}
  {0.2\baselineskip}
  {\sc \centering}
}

\renewcommand{\subsubsection}{
  \@startsection
  {subsubsection}
  {1}
  {0pt}
  {1.1\baselineskip}
  {0.2\baselineskip}
  {\sc \centering}
}

\makeatother

\begin{document}

\title{\large\sc Can robust optimization offer improved portfolio performance?: An empirical study of Indian market}
\normalsize
\author{\sc{Shashank Oberoi} \thanks{Indian Institute of Technology Guwahati, Guwahati-781039, Assam, India, e-mail: s.oberoi@iitg.ac.in}
\and \sc{Mohammed Bilal Girach} \thanks{Indian Institute of Technology Guwahati, Guwahati-781039, Assam, India, e-mail: m.girach@iitg.ac.in}
\and \sc{Siddhartha P. Chakrabarty} \thanks{Indian Institute of Technology Guwahati, Guwahati-781039, Assam, India, e-mail: pratim@iitg.ac.in,
Phone: +91-361-2582606, Fax: +91-361-2582649}}
\date{}
\maketitle
\begin{abstract}

The emergence of robust optimization has been driven primarily by the necessity to address the demerits of the Markowitz model. There has been a noteworthy debate regarding consideration of robust approaches as superior or at par with the Markowitz model, in terms of portfolio performance. In order to address this skepticism, we perform empirical analysis of three robust optimization models, namely the ones based on box, ellipsoidal and separable uncertainty sets. We conclude that robust approaches can be considered as a viable alternative to the Markowitz model, not only in simulated data but also in a real market setup, involving the Indian indices of S\&P BSE 30 and S\&P BSE 100. Finally, we offer qualitative and quantitative justification regarding the practical usefulness of robust optimization approaches from the point of view of number of stocks, sample size and types of data.


\end{abstract}

\section{Introduction}
\label{Introduction}

The risk associated with individual assets can be reduced through investment in a diversified portfolio comprising of several assets. For optimal allocation of weights in a diversified portfolio, one of the well established methods is the classical mean-variance portfolio optimization introduced by Markowitz \cite{Markowitz52,Markowitz59}. Despite being considered as the basic theoretical framework in the field of portfolio optimization, the Markowitz model is not widely accepted among investment practitioners. One of the most major limitations of the mean-variance model is the sensitivity of the optimal portfolios to the errors in the estimation of return and risk parameters. These parameters are estimated using sample mean and sample covariance matrix, which are maximum likelihood estimates (MLEs) (calculated using historical data) under the assumption that the asset returns are normally distributed. According to DeMiguel and Nogales \cite{DeMiguel09}, since the efficiency of MLEs is extremely sensitive to deviations of the distribution of asset returns from the assumed normal distribution, it results in the optimal portfolios being vulnerable to the errors in estimation of input parameters. While referring to the Markowitz model as ``estimation-error maximizers'', Michaud \cite{Michaud89} argues that it often overweights those assets having higher estimated expected return, lower estimated variance of returns and negative correlation between their returns (and vice versa). Best and Grauer \cite{Best91} study the sensitivity of weights of optimal portfolios with respect to changes in estimated expected returns of individual assets. Upon imposition of no short selling constraint, they observe that small changes in estimated expected return of individual assets can result in the assignment of zero weights to almost half the assets comprising the portfolio (which is counterintuitive), leading to a large adjustment in portfolio weights. In an empirical study, Broadie \cite{Broadie93} reports evidence of overestimation of expected returns of optimal portfolios obtained using the Markowitz model by observing that the estimated efficient frontier lies above the actual efficient frontier.

Significant work has been done in the area of robust portfolio optimization in order to address the concerns about the sensitivity of the optimal portfolio to the estimated parameters for the Markowitz model. The robust optimization approach incorporates uncertainty in the input parameters directly into the optimization problem. T{\"u}t{\"u}nc{\"u} and Koenig \cite{Tutuncu04} describe uncertainty using an uncertainty set that includes almost all possible realizations of the uncertain input parameters. The robust portfolio optimization involves optimizing the portfolio performance under the worst possible realizations of the uncertain input parameters. They conduct numerous experiments applying the robust allocation methods to the market data and conclude that robust optimization can be considered as a viable asset allocation alternative for conservative investors. According to Ceria and Stubbs \cite{Ceria06}, the standard approach of robust optimization is too conservative. They argue that it is too pessimistic to adjust the return estimate of each asset downwards. Accordingly, they introduce new variants of robust optimization and observe superior performance of robust approaches vis-\`a-vis the mean-variance analysis, in the majority of cases. Utilizing the standard framework of robust optimization, Scherer \cite{Scherer07} shows that robust methods are equivalent to Bayesian shrinkage estimators and do not lead to significant change in the efficient set. Based on simulations, he observes that robust portfolio underperforms in comparison to the portfolio obtained using the Markowitz model, especially in the case of low risk aversion and high uncertainty aversion. Santos \cite{Santos10} performs similar experiments to compare two types of robust approaches, namely, the standard robust optimization discussed in Scherer's work \cite{Scherer07} and zero net alpha-adjusted robust optimization proposed by Ceria and Stubbs \cite{Ceria06}, with the traditional optimization methods. The empirical results indicate better performance of robust approaches in comparison to the portfolios constructed using mean-variance analysis in the case of simulated data unlike in the case of real market data.

The main aim of the paper is to assess the viability of robust optimization as opposed to the mean-variance optimization, from a practitioner's point of view. In accordance with our motivation, we carry out empirical analysis of the robust models with three uncertainty sets and compare their performance with that of classical Markowitz model, using not only simulated data but also the real market data. We intend to answer various questions related to wider acceptability of robust optimization, both quantitatively and qualitatively from various standpoints. For the purpose of illustration, we have chosen to use the data obtained from the Indian indices of S\&P BSE 30 and S\&P BSE 100.

The rest of the paper is organized as follows. Section \ref{Robust_Portfolio_Optimization_Approaches} discusses the robust portfolio optimization methods used in the work. In section \ref{Computational_Results}, we present the empirical results observed on comparing the performance of robust optimization models with the Markowitz model. Section \ref{Discussion} analyzes the practical usefulness from the point of view of number of stocks and sample size as well as types of data. Finally, we sum up the main takeaways from this work in Section \ref{Conclusion}.

\section{Robust Portfolio Optimization Approaches}
\label{Robust_Portfolio_Optimization_Approaches}

The determination of the structure of uncertainty sets, so as to obtain computationally tractable solutions, is a key step in robust optimization. In the real world, even the distribution of asset returns has an uncertainty associated with it. In order to address this issue, a frequently used technique is to find an estimate of the uncertain parameter and define a geometric bound around it. Empirically, historical data is used to compute estimates of these uncertain parameters. For a given optimization problem, determining the geometry of the uncertainty set is a difficult task. For the purpose of this work, we will use three types of uncertainty sets, namely, box and ellipsoidal (for expected returns) \cite{Fabozzi07,Kim14} and separable (for both expected returns and covariance matrix of returns) \cite{Lu06}. Accordingly, we first introduce the notations to be used in this work.
\begin{enumerate}
\item $N$: Number of assets.
\item $\displaystyle{\mathbf{x}}$: Weight vector for a portfolio.
\item $\displaystyle{\mathbf{a}=\begin{pmatrix} a_{1}, a_{2}, \dots , a_{N} \end{pmatrix}}$: Vector of $N$ uncertain parameters.
\item $\displaystyle{\mathbf{\hat{a}}=\begin{pmatrix}\hat{a}_{1}, \hat{a}_{2}, \dots, \hat{a}_{N} \end{pmatrix}}$: Estimate for $\displaystyle{\mathbf{a}}$.
\item $\displaystyle{\boldsymbol{\mu}}$: Vector for expected return.
\item $\displaystyle{\boldsymbol{{\hat{\mu}}}}$: Estimate for $\displaystyle{\boldsymbol{\mu}}$.
\item $\displaystyle{\Sigma}$: Covariance matrix for asset returns.
\item $\displaystyle{\Sigma_{\mu}}$: Covariance matrix for errors in estimation.
\item $\lambda$: Risk aversion.
\item $\displaystyle{\mathcal{U}_{\mu,\Sigma}}$: General uncertainty set with $\mu$ and $\Sigma$ as uncertain parameters.
\item $\displaystyle{\mathbf{1}}$: Unity vector of length $N$.
\end{enumerate}

The classical Markowitz model formulation with no short selling constraint is given by the following problem (hereafter referred to as \textbf{Mark}):
\begin{equation}
\label{eq:classical_markowitz}
\max\limits_{\mathbf{x}}\left\{\boldsymbol{\mu}^{\top}\mathbf{x}-\lambda\mathbf{x^{\top}}\Sigma\mathbf{x}\right\}~\text{such that}~
\mathbf{x^{\top}}\mathbf{1}=1~\text{and}~\mathbf{x}\geq 0.
\end{equation}

Robust portfolio optimization involves enhancing the robustness of the portfolio obtained using the Markowitz model, by optimizing the portfolio performance in worst-case scenarios. Most of the robust models deal with optimizing a given objective function with a predefined ``uncertainty set'' for obtaining computationally tractable solutions. For any general uncertainty set $\displaystyle{\mathcal{U}_{\mu,\Sigma}}$, the worst case classical Markowitz model formulation \cite{Halldorsson03,Kim14} with no short selling constraint is given as:
\begin{equation}
\label{eq:worst_case_classical_markowitz}
\max\limits_{\mathbf{x}}\left\{\min\limits_{\left(\boldsymbol{\mu},\boldsymbol{\Sigma}\right)~\in~\mathcal{U}_{\mu,\Sigma}}\boldsymbol{\mu}^{\top}\mathbf{x}
-\lambda\mathbf{x^{\top}}\Sigma\mathbf{x}\right\}~\text{such that}~\mathbf{x^{\top}}\mathbf{1}=1~\text{and}~\mathbf{x}~\geq 0,
\end{equation}

\subsection{Robust Portfolio Optimization with Box Uncertainty Set}

A general \textit{polytopic} \cite{Fabozzi07} uncertainty set which resembles a box, is defined as,
\begin{equation}
\label{eqn:box_1}
U_{\boldsymbol{\delta}}(\mathbf{\hat{a}})=\left\{\mathbf{a}:\left|a_{i}-\hat{a}_{i}\right| \leq \delta_{i}, i=1,2,3,\dots,N\right\},
\end{equation}
where $\delta_{i}$ represents the value which determines the confidence interval region for asset $i$. As we intend to model the uncertainty in expected returns $\left(\boldsymbol{\mu}\right)$ using box uncertainty sets, we use,
\begin{equation}
\label{eqn:box_2}
U_{\boldsymbol{\delta}}(\boldsymbol{\hat{\mu}})=\left\{\boldsymbol{\mu}: \left|\mu_{i}-\hat{\mu_{i}}\right|\leq \delta_{i}, i=1,2,3,\dots,N \right\}.
\end{equation}
Accordingly, using (\ref{eqn:box_2}), the max-min robust formulation (\ref{eq:worst_case_classical_markowitz}) reduces to the following maximization problem
(hereafter referred to as \textbf{Box}):
\begin{equation}
\label{eqn:box_markowitz}
\max\limits_{\mathbf{x}}\left\{\boldsymbol{\hat{\mu}}^{\top}\mathbf{x}-\lambda\mathbf{x^{\top}}\Sigma\mathbf{x}-\boldsymbol{\delta}^{\top}|\mathbf{x}|\right\}
~\text{such that}~\mathbf{x^{\top}}\mathbf{1}=1~\text{and}~\mathbf{x} \geq 0.
\end{equation}
While dealing with box uncertainty set, we assume that the returns follow normal distribution. Therefore, we define $\delta_{i}$ for $100(1-\alpha)\%$ confidence level as, \[\displaystyle{\delta_{i}=\sigma_{i} z_{\frac{\alpha}{2}} n^{-\frac{1}{2}}},\]
where $z_{\frac{\alpha}{2}}$ represents the inverse of standard normal distribution, $\sigma_{i}$ is the standard deviation of returns of asset $i$ and $n$ is the number of observations of returns for asset $i$.

\subsection{Robust Portfolio Optimization with Ellipsoidal Uncertainty Set}

In order to capture more information from the data, the consideration of the second moment gives rise to another class of uncertainty sets, namely, ellipsoidal uncertainty sets. The ellipsoidal uncertainty set for expected return $\left(\boldsymbol{\mu}\right)$ is expressed as:
\begin{equation}
\label{eqn:ellipsoidal}
U_{\delta}(\boldsymbol{\hat{\mu}})=\left\{\boldsymbol{\mu}: (\boldsymbol{\mu}-\boldsymbol{\hat{\mu}})^{\top}\Sigma^{-1}_{\boldsymbol{\mu}}
(\boldsymbol{\mu}-\boldsymbol{\hat{\mu}})\leq\delta^2 \right\}.
\end{equation}
Therefore, the max-min robust formulation (\ref{eq:worst_case_classical_markowitz}) in conjunction with (\ref{eqn:ellipsoidal}) results in the following maximization problem (hereafter referred to as \textbf{Ellip}):
\begin{equation}
\label{eqn:ellipsoidal_markowitz}
\max\limits_{\mathbf{x}}\left\{\boldsymbol{\hat{\mu}}^{\top}\mathbf{x}-\lambda \mathbf{x}^{\top}\Sigma\mathbf{x}
-\delta\sqrt{\mathbf{x}^{\top}\Sigma_{\boldsymbol{\mu}}\mathbf{x}}\right\}~\text{such that}~\mathbf{x^{\top}}\mathbf{1}=1~\text{and}~\mathbf{x}\geq 0.
\end{equation}
If the uncertainty set follows ellipsoid model, the condidence level is set using a chi-square ($\chi^{2}$) distribution with the number of assets being the degrees of freedom (df).
Accordingly, for $100(1-\alpha)\%$ confidence level, $\delta$ is defined as \cite{Ceria06,Scherer07}:
\begin{equation}
\delta^2=\chi_{N}^2(\alpha)
\end{equation}
where $\chi_{N}^2(\alpha)$ is the inverse of a chi square distribution with $N$ degrees of freedom.

\subsection{Robust Portfolio Optimization with Separable Uncertainty Set}

The above two robust approaches model only the expected returns using uncertainty sets. Hence, in order to also encapsulate the uncertainly in the covariances, the box uncertainty set for the covariance matrix of returns is defined akin to that for expected returns. The lower bound $\underline{\Sigma}_{ij}$ and the upper bound $\overline{\Sigma}_{ij}$ can be specified for each entry $\Sigma_{ij}$, resulting in the following constructed box uncertainty set for the covariance matrix \cite{Tutuncu04}:
\begin{equation}
\label{eqn:separable_1}
U_{\Sigma}=\{\Sigma: \underline{\Sigma} \leq \Sigma \leq \overline{\Sigma},~\Sigma \succeq 0\}.
\end{equation}
In the above equation, the condition $\Sigma \succeq 0$ implies that $\Sigma$ is a symmetric positive semidefinite matrix. T{\"u}t{\"u}nc{\"u} and Koenig \cite{Tutuncu04} define the uncertainty set for expected returns as,
\begin{equation}
\label{eqn:separable_2}
U_{\boldsymbol{\mu}}=\{\boldsymbol{\mu}:\underline{\boldsymbol{\mu}}\leq\boldsymbol{\mu}\leq\overline{\boldsymbol{\mu}}\},
\end{equation}
where $\underline{\boldsymbol{\mu}}$ and $\overline{\boldsymbol{\mu}}$ represent lower and upper bounds on expected return vector $\boldsymbol{\mu}$ respectively. Consequently, the max-min robust formulation (\ref{eq:worst_case_classical_markowitz}) transforms to the following maximization problem (hereafter referred to as \textbf{Sep}):
\begin{equation}
\label{eqn:separable_markowitz}
\max_{\mathbf{x}} \left\{\underline{\boldsymbol{\mu}}^{\top}\mathbf{x}-\lambda \mathbf{x^{\top}}\overline{\Sigma}\mathbf{x}\right\}~\text{such that}~ \mathbf{x^{\top}}\mathbf{1}=1~\text{and}~\mathbf{x}\geq 0.
\end{equation}
The above approach involves the use of ``separable'' uncertainty sets \cite{Lu06}, which implies that the uncertainty sets for expected returns and covariance matrix are defined independent of each other.

\section{Computational Results}
\label{Computational_Results}

In this section, we analyze the performance of the robust portfolio optimization approaches discussed in Section \ref{Robust_Portfolio_Optimization_Approaches} vis-\`a-vis the Markowitz model, using the historical data from the market, as well as simulated data. For the purpose of this analysis, we consider two scenarios in terms of number of stocks $N$, being $31$ and $98$, with the goal of observing the effect of increase in the number of stocks on the performance of robust portfolio optimization approaches. These numbers were chosen since they represent the number of stocks in S\&P BSE 30 and S\&P BSE 100 indices, respectively.

For the first scenario ($N=31$), we make use of the daily log-returns, based on the adjusted daily closing prices of the $31$ stocks comprising the S\&P BSE 30, obtained from Yahoo Finance \cite{yf}. Accordingly, we consider the period of our analysis to be from 18th December, 2017 to 30th September, 2018 (both inclusive) which had a total of $194$ active trading days \textit{i.e.,} $193$ daily log-returns. Corresponding to this historical data from S\&P BSE 30, we generate two sets of simulated data for all the $31$ assets, by sampling returns using a multivariate normal distribution whose mean and covariance matrix are set to those obtained for the historical S\&P BSE 30 data. The first set of sample returns comprises of the number of samples to be the same as in the historical data, namely $193$, in this case. On the other hand, the second set comprises of a larger number of samples, namely, $1000$. The two sets of simulated sample returns of different sizes were used to facilitate the study of the impact of the number of samples in simulated data on the performance of the robust portfolio optimization approaches. We make a comparative study of robust portfolio optimization approaches, in case of the historical S\&P BSE 30 data, as well as the two sets of simulated data, in order to analyze whether the worst case robust portfolio optimization approaches are useful in a real market setup. For the second scenario ($N=98$), we use the daily log-returns, based on the adjusted daily closing prices of the $98$ stocks comprising the S\&P BSE 100, obtained from Yahoo Finance \cite{yf}, for the period 18th December, 2016 to 30th September, 2018 (both inclusive) with $443$ trading days \textit{i.e.,} $442$ daily log-returns. The two sets of simulated data were generated in a manner akin to the scenario of $N=31$ assets. Similar kind of comparative study is performed for the second scenario.

For Box and Ellip model, we construct uncertainty sets in expected return with $100(1-\alpha)\%$ confidence level by considering $\alpha=0.05$. Separable uncertainty set in Sep model is constructed as a $100(1-\alpha)\%$ confidence interval for both $\mu$ and $\Sigma$ using non-parametric Boostrap Algorithm with same $\alpha$ as in other robust models and assuming $\beta$, \textit{i.e} the number of simulations, equal to $8000$.

The performance analysis for these robust portfolio models vis-\`a-vis the Mark model is performed using the \textit{Sharpe Ratio} of the constructed portfolios, with $\lambda$ representing the risk-aversion in the ideal range \textit{i.e.,} $\lambda \in [2,4]$ \cite{Fabozzi07}. Further, since the T-bill rate in India from 2016 to 2018 was observed to oscillating around $6\%$ \cite{rbi}, so we have assumed the annualized riskfree rate to be equal to $6\%$. We now present the computational results observed in case of the two scenarios, as discussed above.

\subsection{Performance with $N=31$ assets}

\begin{figure}[h]
\centering
\begin{subfigure}{.5\textwidth}
  \centering
  \includegraphics[width=.8\linewidth]{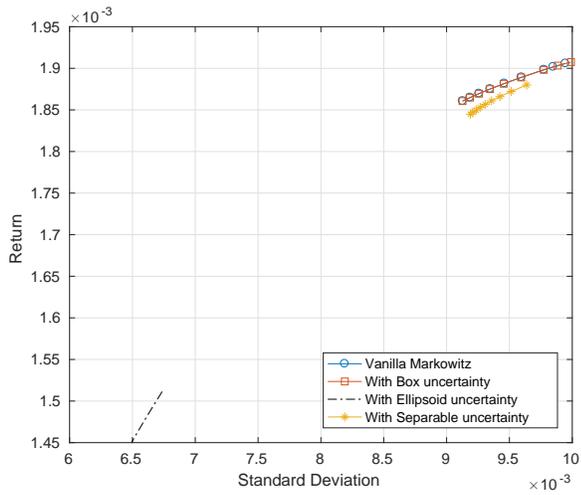}
\end{subfigure}%
\begin{subfigure}{.5\textwidth}
  \centering
  \includegraphics[width=.8\linewidth]{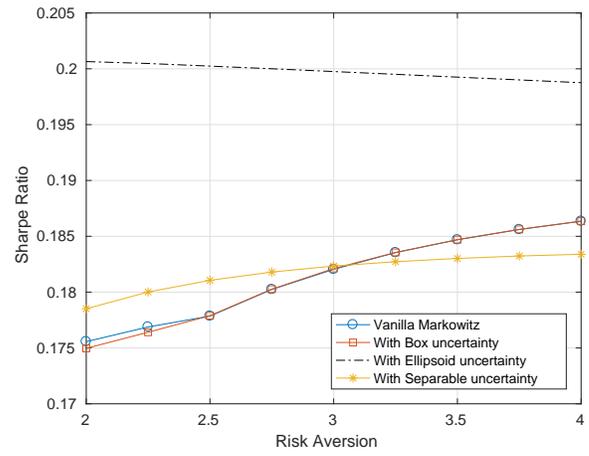}
\end{subfigure}
\caption{Efficient Frontier plot and Sharpe Ratio plot for different portfolio optimization models in case of Simulated Data with 1000 samples (31 assets)}
\label{fig:1}
\end{figure}

We begin with the analysis for $N=31$ assets, in the case of the simulated data with $1000$ samples and present the results in Figure \ref{fig:1} and Table \ref{tab:1}. From Figure \ref{fig:1} we observe that the efficient frontiers for the Ellip and the Sep models lie below the efficient frontier for the Mark model, which  supports the argument made in \cite{Broadie93} regarding over-estimation of the efficient frontier for the Mark model. Further, the observed overlap of the efficient frontiers for the Mark and the Box models suggest that the utilizing box uncertainty sets for robust portfolio optimization does not prove to be of much use in this case. Also, from Figure \ref{fig:1}, we observe that the Mark model starts outperforming the Sep model in terms of the Sharpe Ratio after the risk-aversion $\lambda$ crosses $3$. The above observations are supported quantitatively by the results tabulated in Table \ref{tab:1} as well, since the average Sharpe Ratio for portfolios constructed in the ideal range of risk-aversion $\lambda\in [2,4]$ is the same in case of both the Mark and the Box models. Also, we infer from Table \ref{tab:1}, that the Sep model performs at par with the Mark model by taking the average Sharpe Ratio into consideration, with the Ellip model performing the best among all the models.

\begin{figure}[h]
\centering
\begin{subfigure}{.5\textwidth}
  \centering
  \includegraphics[width=.8\linewidth]{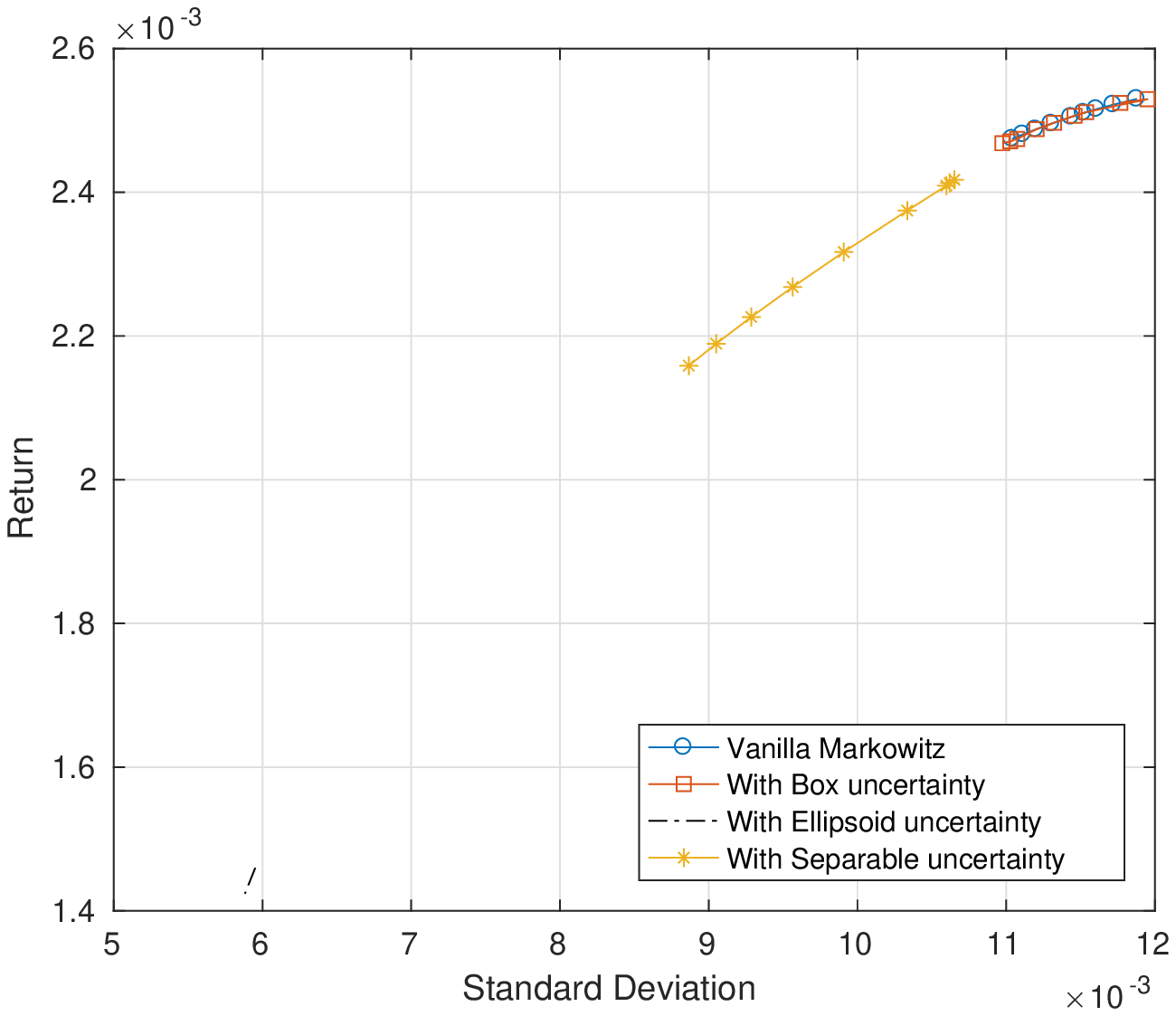}
\end{subfigure}%
\begin{subfigure}{.5\textwidth}
  \centering
  \includegraphics[width=.8\linewidth]{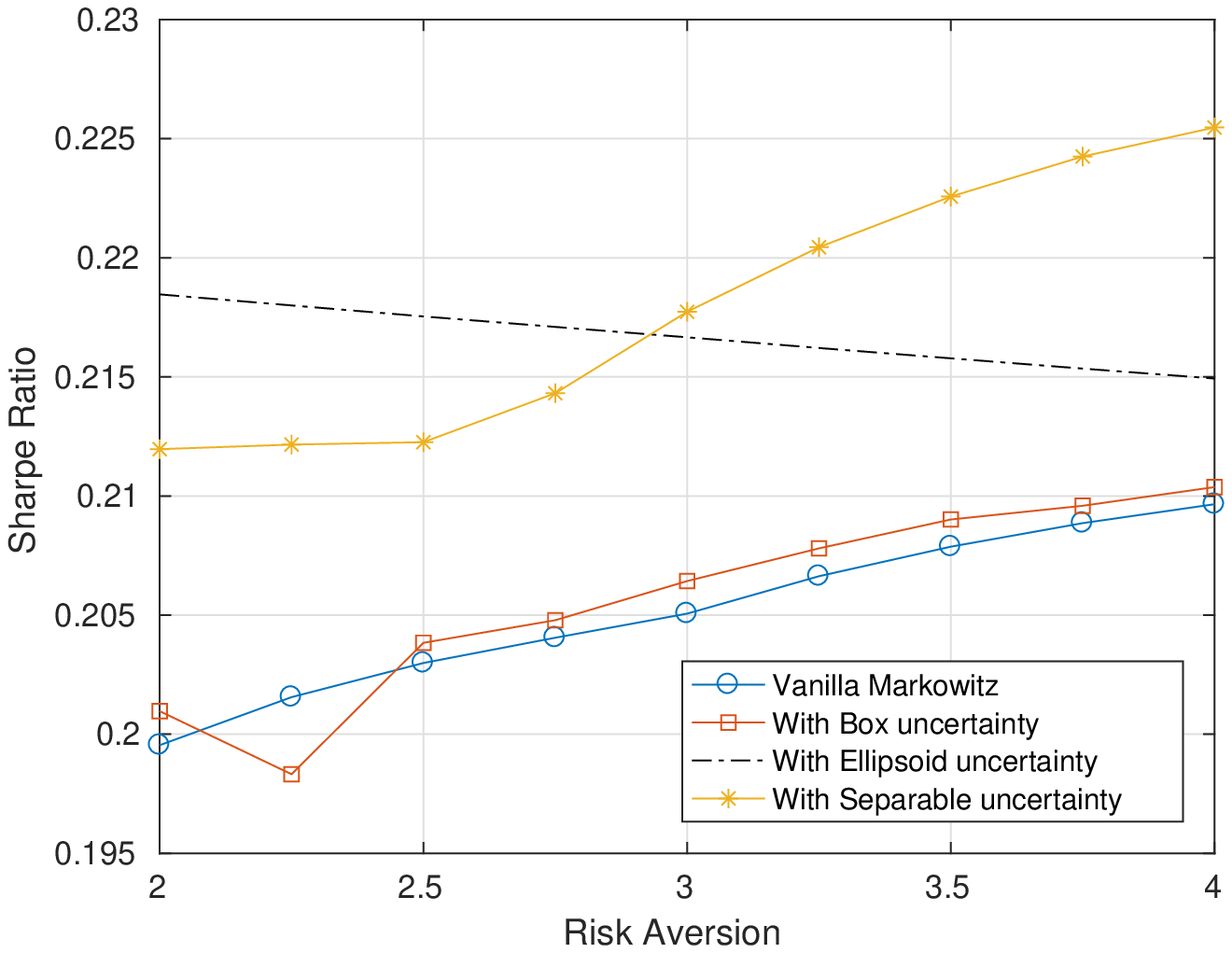}
\end{subfigure}
\caption{Efficient Frontier plot and Sharpe Ratio plot for different portfolio optimization models in case of Simulated Data with same number of samples as market data (31 assets)}
\label{fig:2}
\end{figure}

The analysis with the number of simulated samples being the same as the number of log-returns in the case of S\&P BSE 30 data is presented in Figure \ref{fig:2} and Table \ref{tab:2}. The efficient frontiers for the Sep and Ellip model lie below that of the Mark model. We observe results similar to the case when $1000$ simulated samples were considered, upon comparison of the Mark model and the Box model. However, we observe a slight inconsistency in the performance of Box model as evident from the plot of the Sharpe Ratio in Figure \ref{fig:2}.  We also infer that the Ellip model and the Sep model outperform the Mark model in terms of the Sharpe Ratio in the ideal range of risk-aversion $\lambda \in [2,4]$. It is difficult to compare the performance of the Ellip model with that of the Sep model in this case, since the average Sharpe Ratio for both of them is almost the same (Table \ref{tab:2}).

\begin{figure}[h]
\centering
\begin{subfigure}{.5\textwidth}
  \centering
  \includegraphics[width=.8\linewidth]{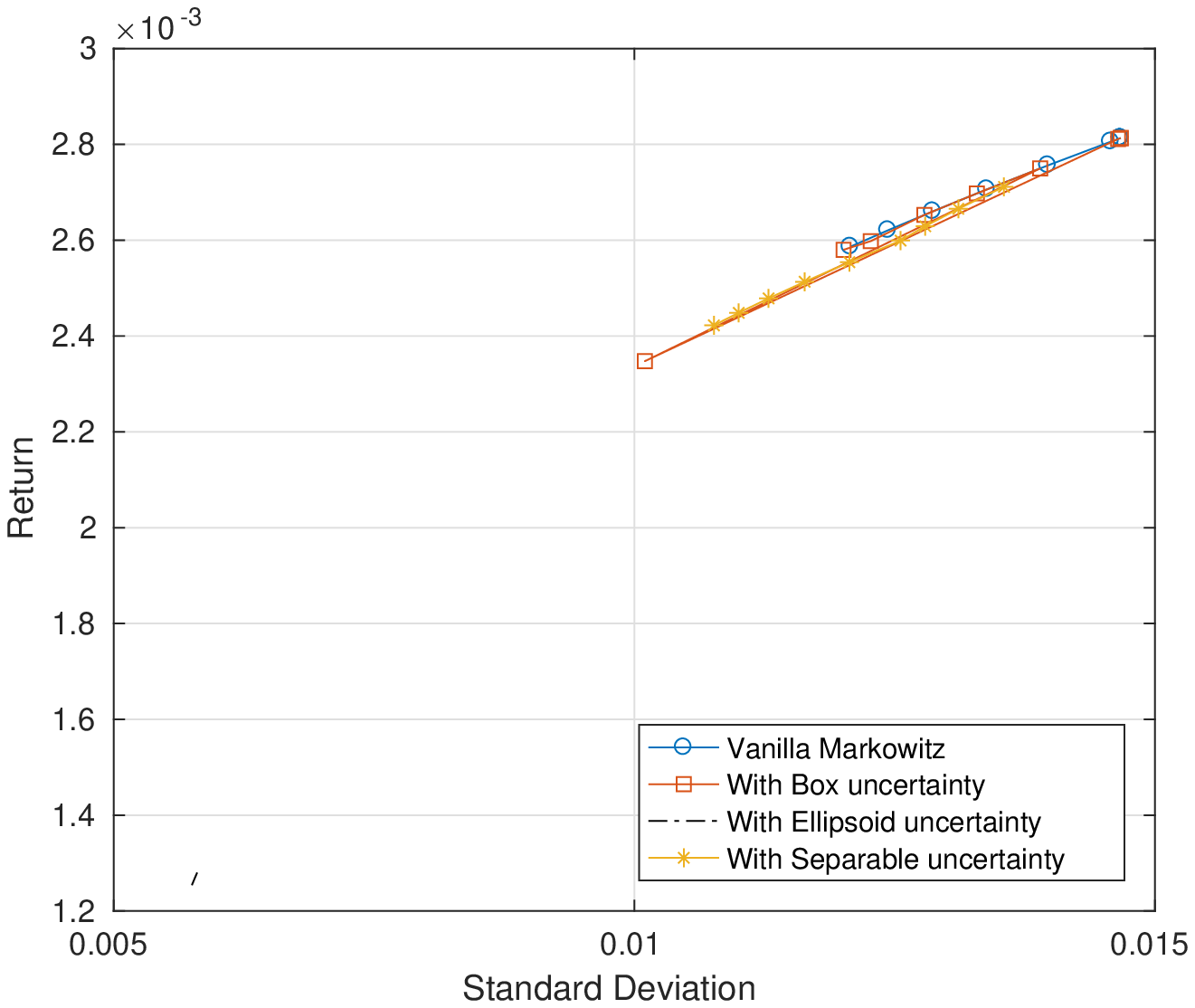}
\end{subfigure}%
\begin{subfigure}{.5\textwidth}
  \centering
  \includegraphics[width=.8\linewidth]{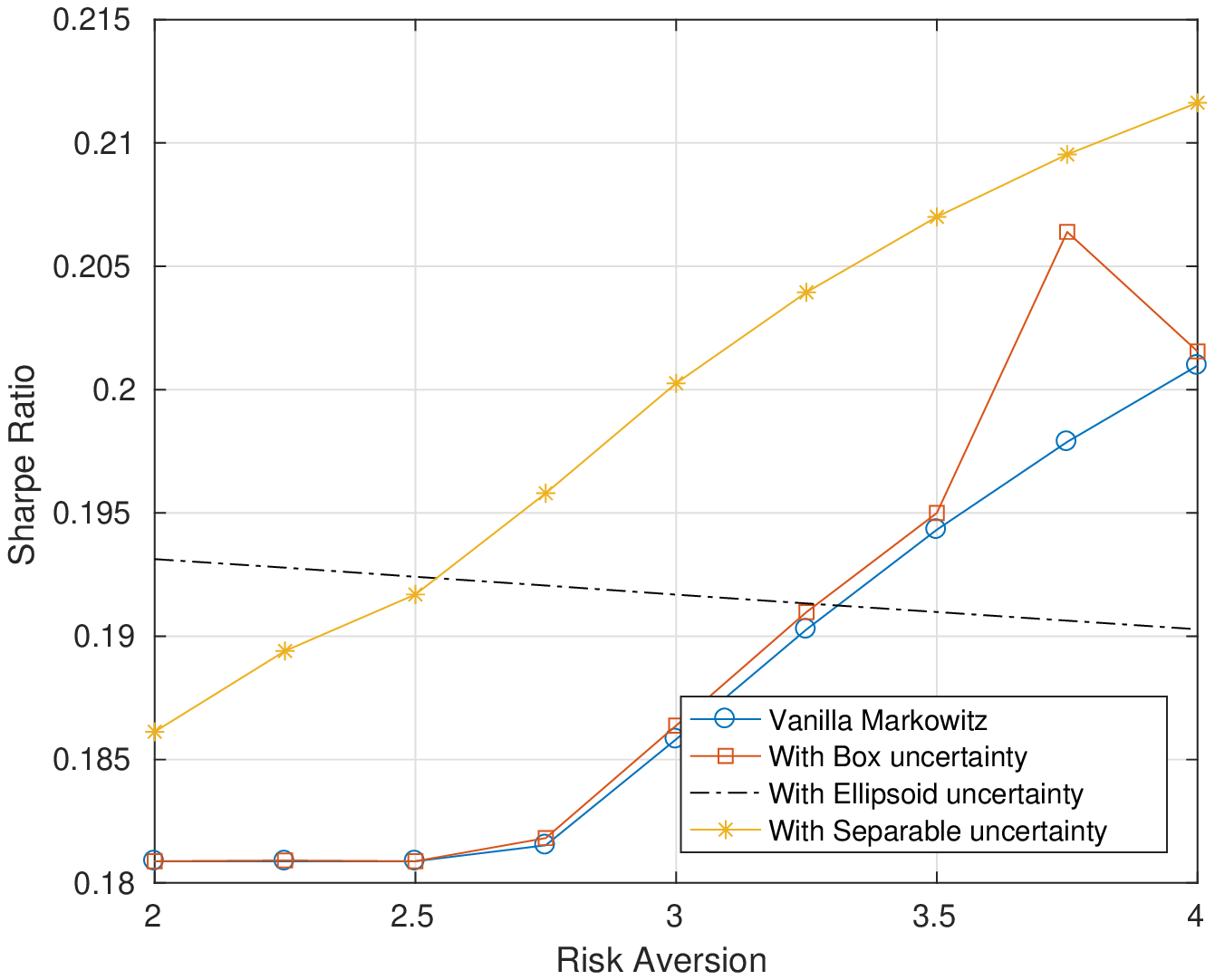}
\end{subfigure}
\caption{Efficient Frontier plot and Sharpe Ratio plot for different portfolio optimization models in case of Market Data (31 assets)}
\label{fig:3}
\end{figure}

For the historical market data involving the stocks comprising S\&P BSE 30, we observe from Figure \ref{fig:3}, that the efficient frontiers for the Mark model and the Box model almost overlap with each other. Further, the efficient frontier for the Sep model lies below that of the Mark model with further widening of the gap between the plots, in case of the Ellip model. However, the performance of the Box model, in terms of the Sharpe ratio is quite inconsistent as evident from Figure \ref{fig:3}. We also observe that the Sep model outperforms the Mark model in the ideal range of risk-aversion $\lambda\in [2,4]$ upon taking the Sharpe Ratio into consideration as the performance measure. This is not true in case of the Ellip Model, as evident from the Sharpe Ratio plot in Figure \ref{fig:3}. Even from Table \ref{tab:3}, we observe that average Sharpe Ratio for the Ellip model is only slightly greater than that for the Mark model. We also note that the Sep model outperforms all the other three models.

\textit{A common observation that could be inferred from three cases considered in the scenario involving less number of assets ($N=31$) is that the Sep and the Ellip models perform superior or equivalent in comparison to the Mark model in the ideal range of risk-aversion.}

\subsection{Performance with $N=98$ assets}

\begin{figure}[h]
\centering
\begin{subfigure}{.5\textwidth}
  \centering
  \includegraphics[width=.8\linewidth]{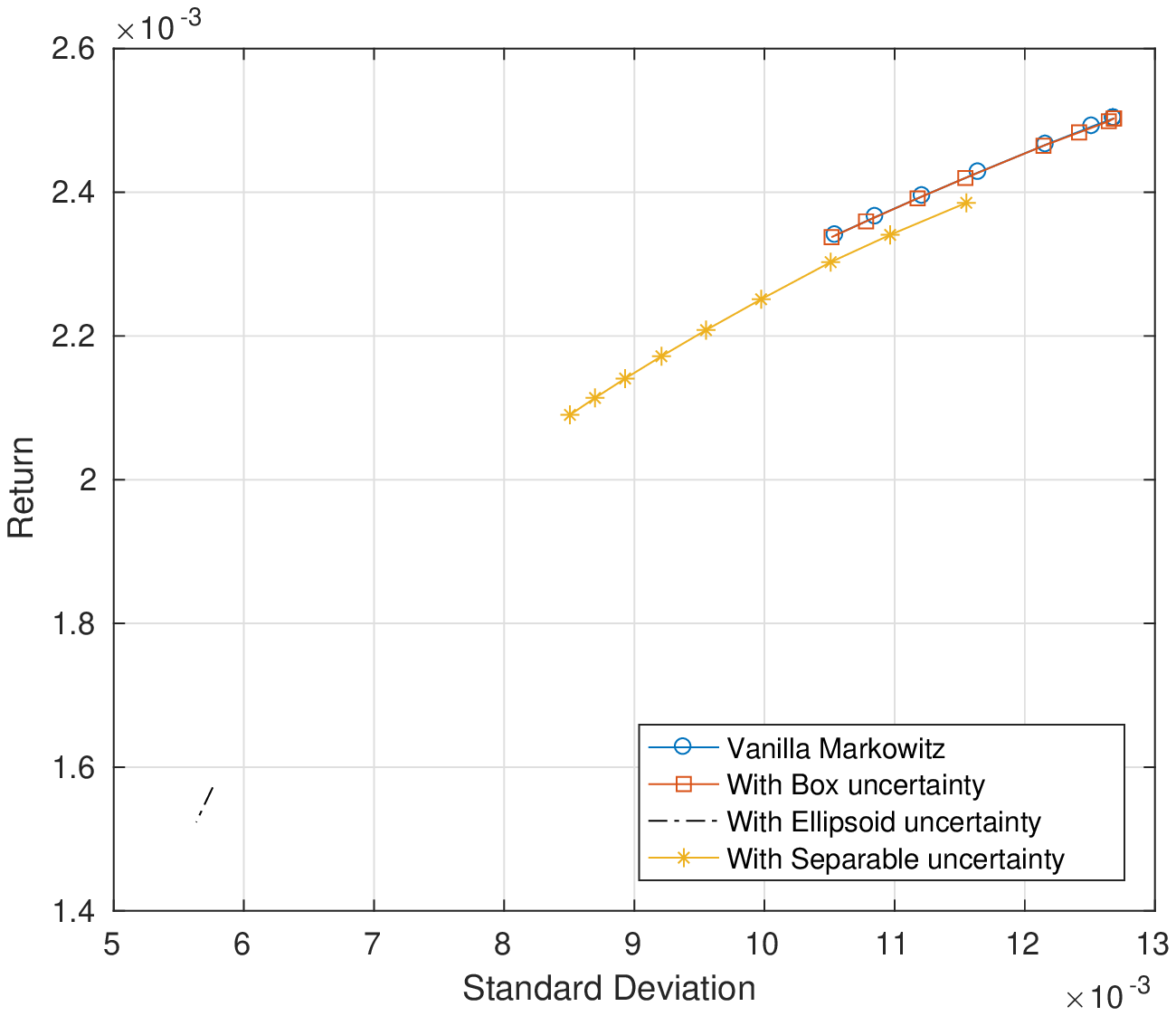}
\end{subfigure}%
\begin{subfigure}{.5\textwidth}
  \centering
  \includegraphics[width=.8\linewidth]{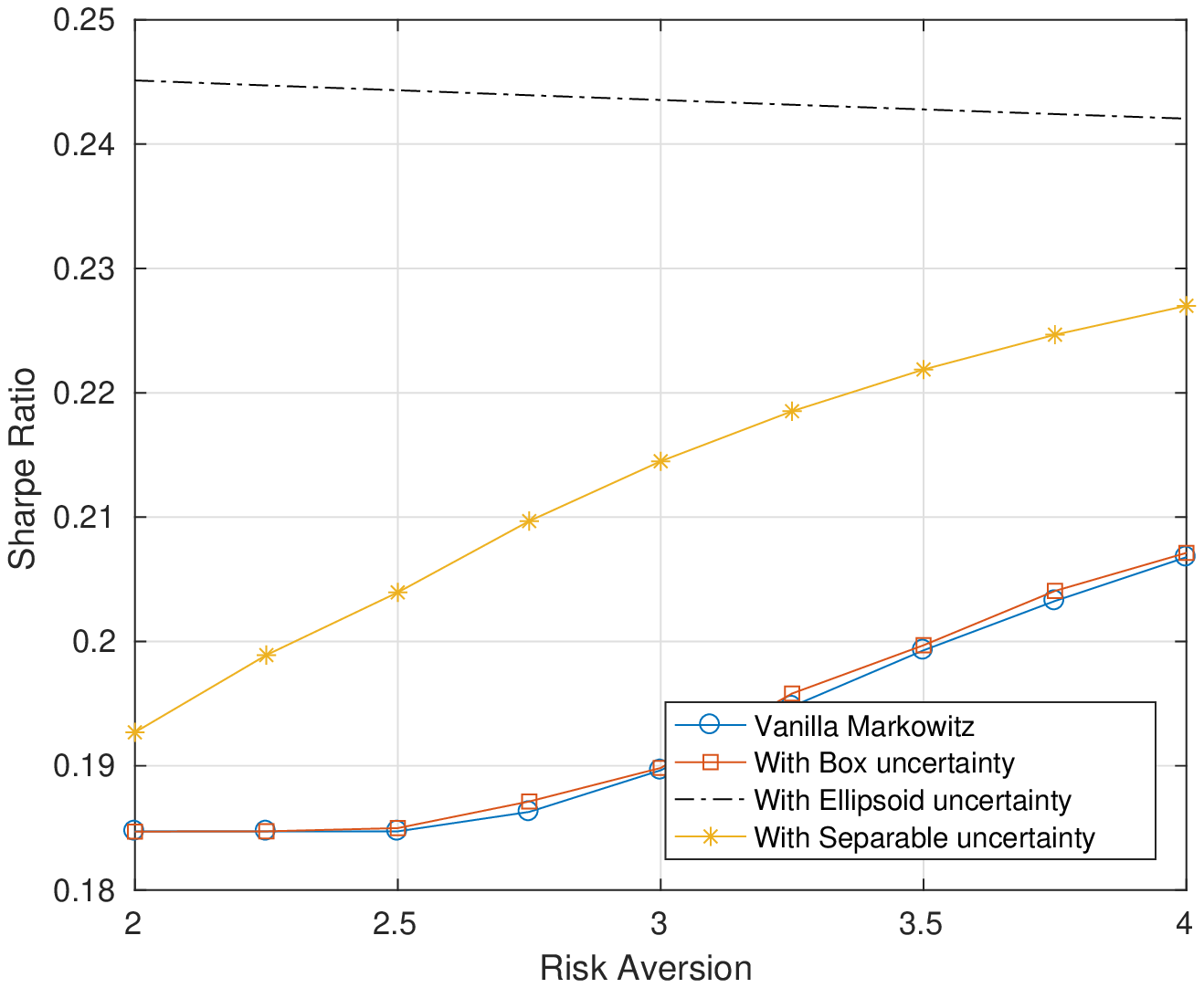}
\end{subfigure}
\caption{Efficient Frontier plot and Sharpe Ratio plot for different portfolio optimization models in case of Simulated Data with 1000 samples (98 assets)}
\label{fig:4}
\end{figure}

We now analyze the scenario involving $N=98$ assets. On applying robust model along with the Mark model on the simulated data having $1000$ samples, we observe results similar to the corresponding case for the previous scenario when we compared the Box model with the Mark model. This is evident from the coinciding plots of the efficient frontier and the plots for the Sharpe Ratio for both the models in Figure \ref{fig:4}. However, in contrast to the scenario of $N=31$ assets, we observe that not only does the Ellip model but also the Sep model outperforms the Mark model when considering the portfolios constructed in the ideal range of risk-aversion $\lambda\in [2,4]$. Additionally, from Table \ref{tab:4}, we can infer that the Ellip model exhibits superior performance in comparison to the Sep model, in terms of greater average value of the Sharpe Ratio.

\begin{figure}[h]
\centering
\begin{subfigure}{.5\textwidth}
  \centering
  \includegraphics[width=.8\linewidth]{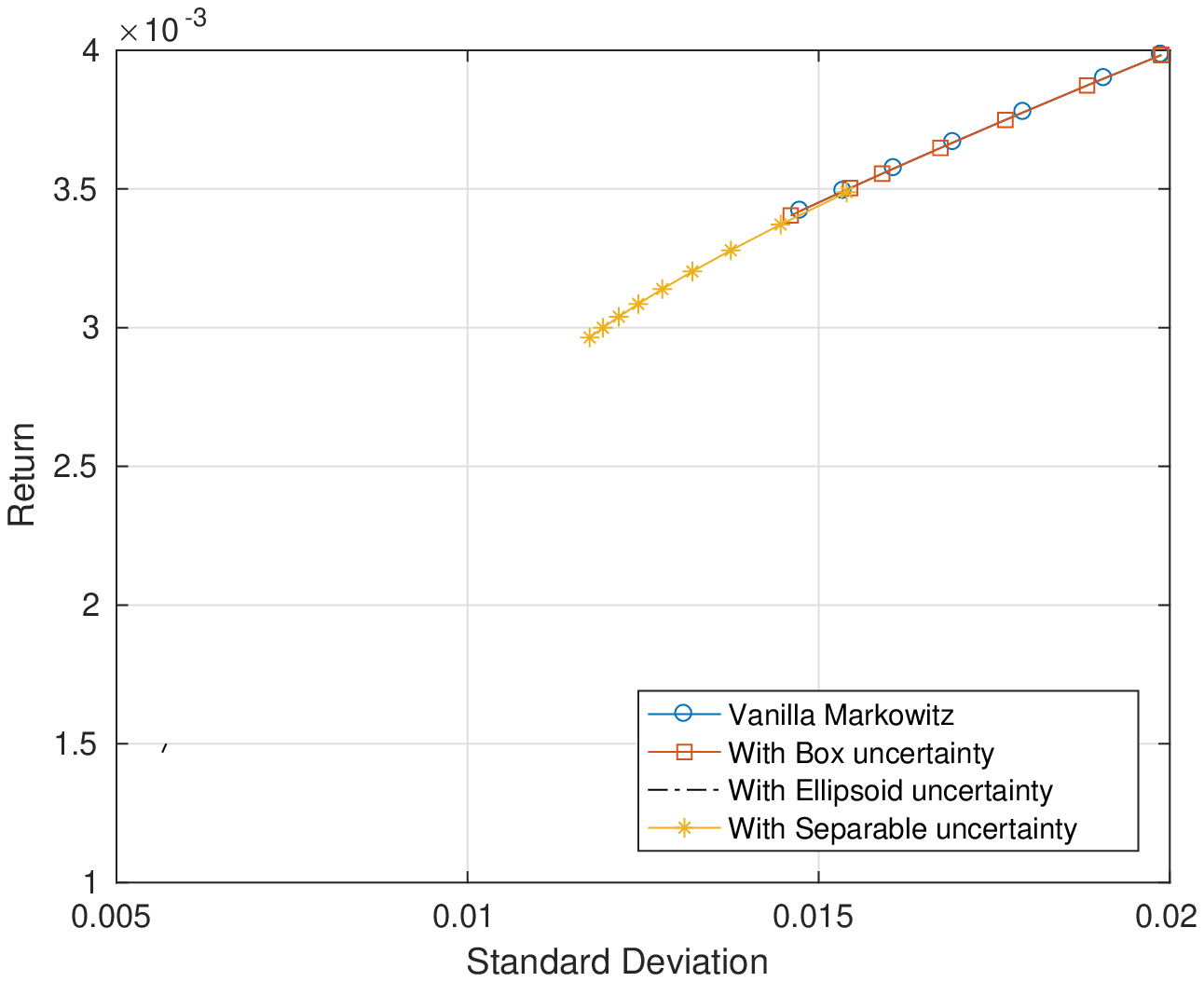}
\end{subfigure}%
\begin{subfigure}{.5\textwidth}
  \centering
  \includegraphics[width=.8\linewidth]{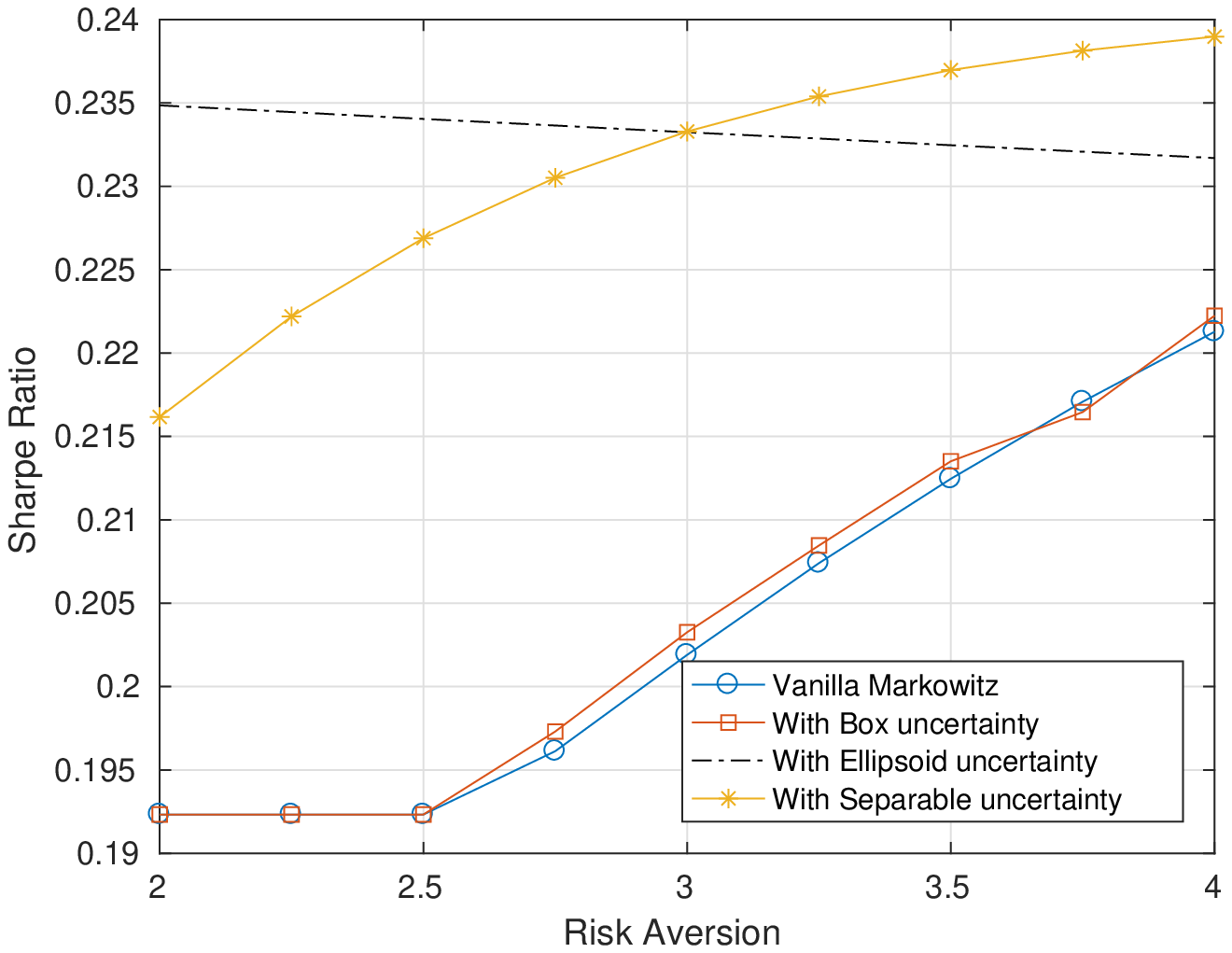}
\end{subfigure}
\caption{Efficient Frontier plot and Sharpe Ratio plot for different portfolio optimization models in case of Simulated Data with same number of samples as market data (98 assets)}
\label{fig:5}
\end{figure}

In Figure \ref{fig:5} and Table \ref{tab:5} we present the results of the study for the simulated data with the number of samples being the same as that of log-returns of S\&P BSE 100 data. The comparative results observed for the Box model and the Mark model are similar to the previous case of $1000$ simulated samples. In the ideal range of risk aversion $\lambda\in [2,4]$, one observes that the efficient frontier for both the Ellip as well as the Sep model lie below the efficient frontier for the Mark model and both the models perform better than the Mark model in terms of the Sharpe Ratio. Additionally, from the Sharpe ratio plot in Figure \ref{fig:5}, any comparative inference of the Sep model and the Ellip model is difficult, since each outperforms the other in a different sub-interval of the risk-aversion range. The similar values of the average Sharpe Ratio in Table \ref{tab:5} supports the claim of almost equivalent performance of these two models in this case.

\begin{figure}[h]
\centering
\begin{subfigure}{.5\textwidth}
  \centering
  \includegraphics[width=.8\linewidth]{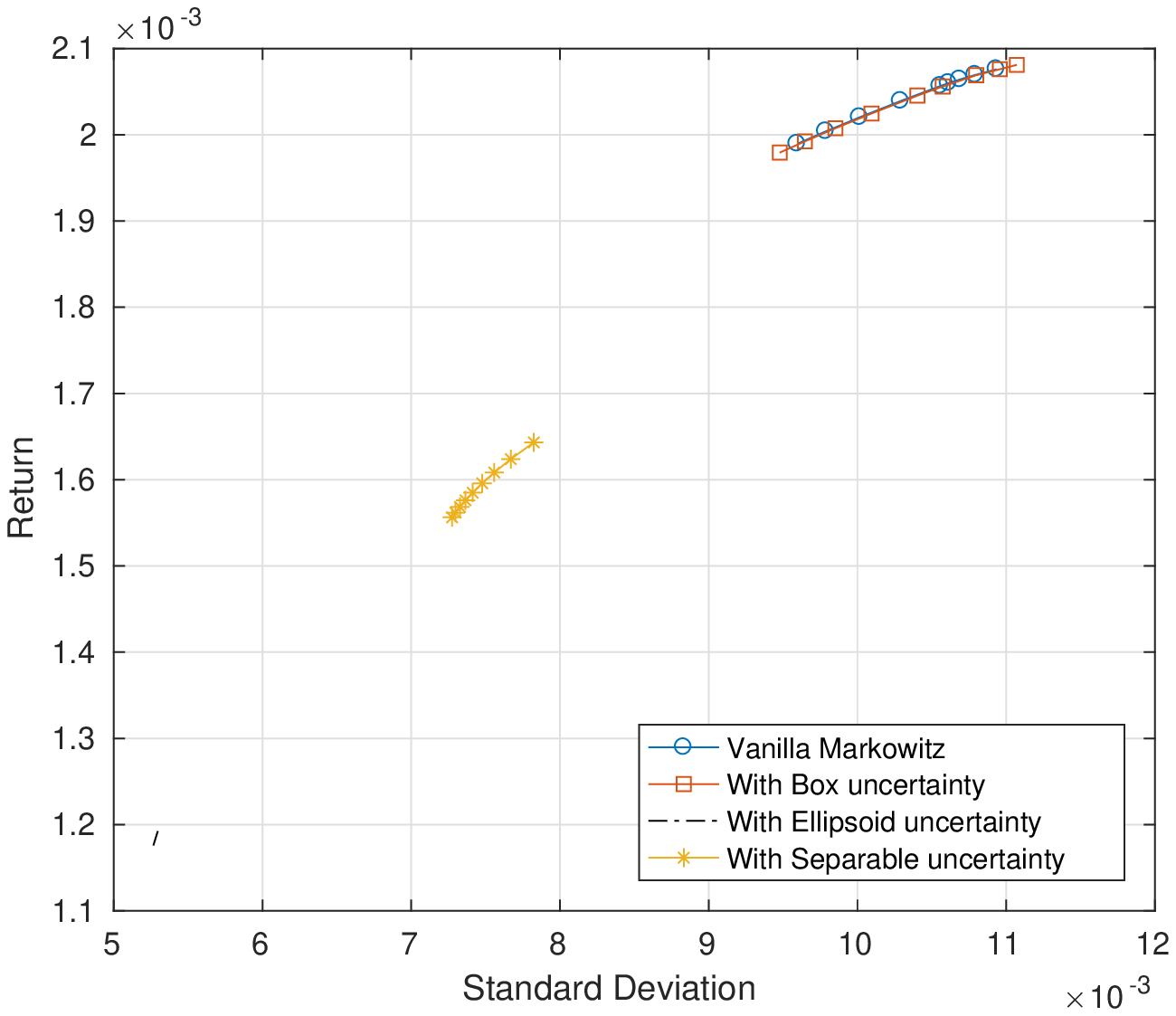}
\end{subfigure}%
\begin{subfigure}{.5\textwidth}
  \centering
  \includegraphics[width=.8\linewidth]{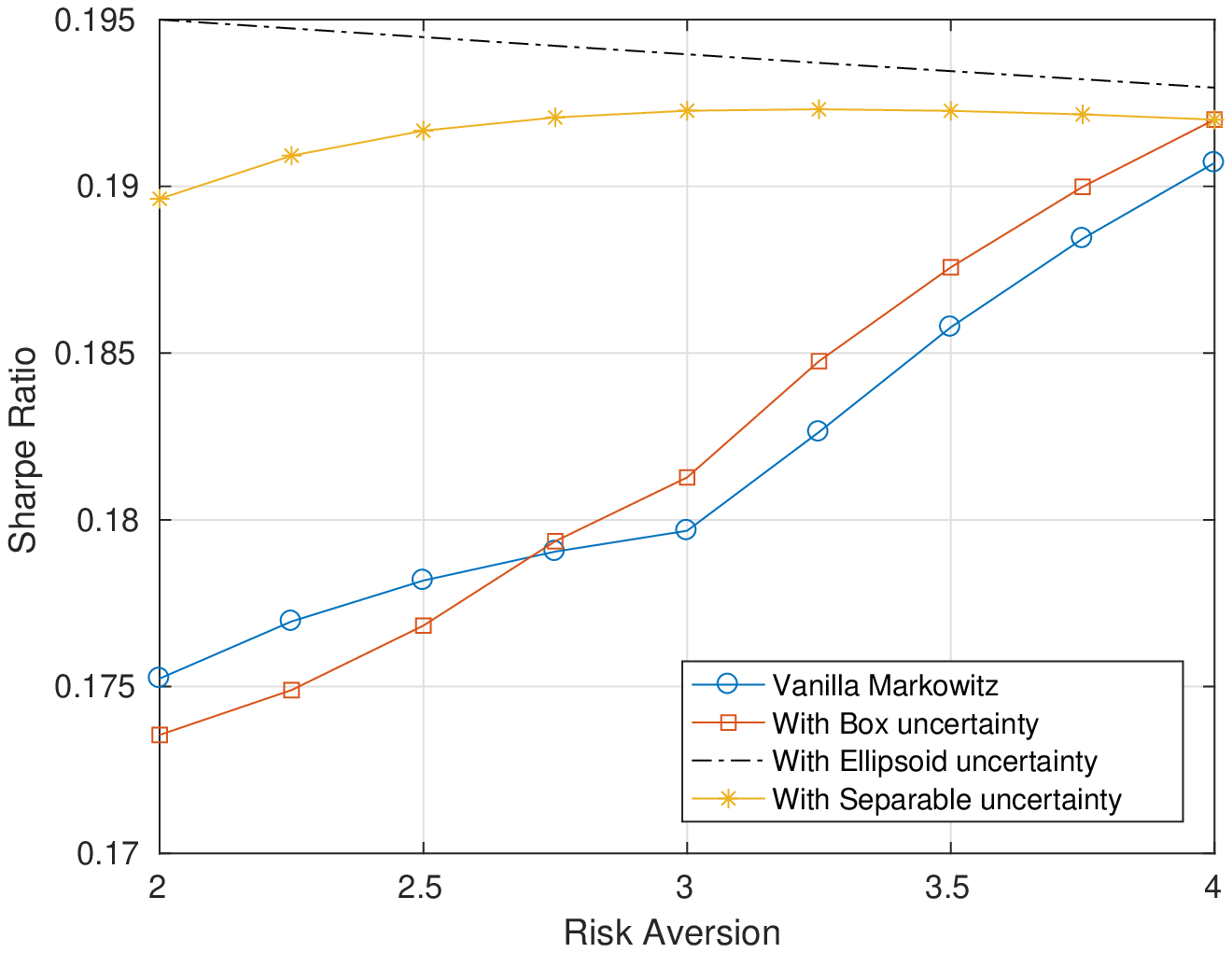}
\end{subfigure}
\caption{Efficient Frontier plot and Sharpe ratio plot for different portfolio optimization models in case of Market Data (98 assets)}
\label{fig:6}
\end{figure}

Finally, the results for the historical market data, involving stocks comprising S\&P BSE 100 are presented in Figure \ref{fig:6} and Table \ref{tab:6}. While the efficient frontier plot leads to observations similar to the previous case, however, there is a slight inconsistency in the performance of the Box model as can be seen from the plot of the Sharpe Ratio in Figure \ref{fig:6}. The robust portfolios constructed using the Sep and the Ellip models outperform the ones constructed using the Mark model in the ideal range of risk-aversion $\lambda\in [2,4]$. Additionally, the performance of the Ellip model is marginally better than the Sep model as evident from the Sharpe Ratio plot, an inference that is supported by the marginal difference in average Sharpe Ratio of both (Table \ref{tab:6}).

\textit{We draw a common inference from the three cases considered in the scenario involving greater number of assets, \textit{i.e.}, the Sep and the Ellip model outperform the Mark model in the ideal range of risk aversion.}

\section{Discussion}
\label{Discussion}

In this concluding section we analyze the different kinds of scenarios in the context of trends of the Sharpe Ratio. Recall that, we have considered the ``adjusted closing prices'' data of S\&P BSE 30 and S\&P BSE 100 to illustrate our analysis. Further, we have also generated simulated samples using the true mean and covariance matrix of log-returns obtained from the aforesaid actual market data of ``adjusted closing prices''. Since the number of instances in market data for the assets comprising the two indices, was very less, we simulated two sets of samples, one where the number of simulated samples matches the number of instances of real market data available, say $\zeta(<1000)$ and another where the number of simulated samples is large (a constant, which in our case was taken to be $1000$), irrespective of the number of stocks. The motivation behind this setup was to understand if the market data we obtained (which was limited) is able to capture the trends and results in better portfolio performance.

\subsection{From the Standpoint of Number of Stocks}

We begin with a description of the results summarized in Table \ref{tab:no_stocks}, wherein for a particular row and a particular column, we presented the maximum possible Sharpe Ratio that was obtained for that particular scenario. For example, in case of the tabular entry for the case when $N=98$ where we simulated $\zeta$ samples using true mean vector and the true covariance matrix of S\&P BSE 100, we refer to Table \ref{tab:5} (which explains the simulation corresponding to S\&P BSE 100 with $\zeta$ simulated samples) and take the maximum of its last row \textit{i.e.}, maximum of average Sharpe ratios that was attained using the available robust and Mark models.

Larger the number of stocks, better is the performance of the portfolios constructed using robust optimization. This claim can be supported both qualitatively and quantitatively. Qualitatively, the number of stocks in a portfolio represents its diversification. According to Modern Portfolio Theory (MPT), investors get the benefit of better performance from diversifying their portfolios since it reduces the risk of relying on only one (small number) asset (assets) to generate returns. Based on the analysis by Value Research Online \cite{vro} one observes that on an average basis, the large-cap funds hold around $38$ shares while the mid-cap funds hold around $50-52$ assets for balanced funds, in which around $65-70\%$ of the assets are held in equity. This is because of great stability of returns in case of companies with large market capitalization, whereas this is not the case with mid-cap companies. Hence diversification requirements drives greater percentages in equities in case of mid-cap funds. From Table \ref{tab:no_stocks}, we can provide quantitative justification by observing that the Sharpe Ratio was more for portfolios with larger number of stocks as compared to portfolios with smaller number of stocks. However, we observe opposite behavior for the market data which can be attributed to the following two reasons:
\begin{enumerate}
\item The insufficient availability of market data, when it comes to larger number of stocks.
\item The error in the estimation of return and covariance matrix accumulating as the number of stocks increases, impacting the performance of the model \cite{Michaud89}.
\end{enumerate}

\subsection{From the Standpoint of Number of Simulated Samples}

We now focus on the performance of the portfolio when different number of samples were simulated and tabulate the results in Table \ref{tab:no_samples} in the same way as was done in the preceding discussion. Here several interesting performance trends can be noticed. We observe that in the case of smaller number of stocks, the performance when the number of simulated samples is $\zeta (< 1000)$ is better than the case when a large ($1000$) simulated samples were generated. On the other hand, the exactly opposite trend can be observed when higher number of stocks are taken into consideration. This observation can be explained as follows: In the case of real market, the number of data instances being available is relatively low. So, when larger number of samples were generated, we observe higher Sharpe Ratio as compared to $\zeta$ number of simulations. However, the reason behind such a pattern of opposite behavior, when smaller number of stocks are considered, is not obvious.

\subsection{From the Standpoint of the Kind of Data}

Finally, we discuss about the performance of the portfolio from the standpoint of kind of data that we have used in this work. Accordingly, the relevant results are tabulated in Table \ref{tab:data_type}, from where the behavior is observed to be fairly consistent. For both the cases, the performance in case of the simulated data is better than in case of the real market data. This is clear from the fact that the real market data is difficult to model as it hardly follows any distribution, whereas the simulated is generated from multivariate normal distribution with mean and covariances as the true values obtained from the data.

\section{Conclusion}
\label{Conclusion}

Robust optimization is an emerging area of portfolio optimization. Various questions have been raised on the advantages of robust methods over the Markowitz model. Through computational analysis of various robust optimization approaches followed by a discussion from different standpoints, we try to address this skepticism. We observe that robust optimization with ellipsoidal uncertainty set performs superior or equivalent as compared to the Markowitz model, in the case of simulated data, similar to the results reported by Santos \cite{Santos10}. In addition, we observe favorable results in the case of market data as well. Better performance of the robust formulation having separable uncertainty set in comparison to the Markowitz model is in line with the previous study on the same robust model by T{\"u}t{\"u}nc{\"u} and Koenig \cite{Tutuncu04}. Empirical results presented in this work advocate enhanced practical use of the robust models involving ellipsoidal uncertainty set and separable uncertainty set and accordingly, these models can be regarded as possible alternatives to the classical mean-variance analysis in a practical setup.

\newpage

\newpage

\begin{table}[h]
\centering
\captionsetup{justification=centering}
\begin{tabular}{||c|c|c|c|c||}
\hline
$\lambda$ & $SR_{Mark}$ & $SR_{Box}$ & $SR_{Ellip}$ & $SR_{Sep}$ \\
\hline
2 & 0.176 & 0.175 & 0.201 & 0.178 \\
2.5 & 0.178 & 0.178 & 0.2 & 0.181 \\
3 & 0.182 & 0.182 & 0.2 & 0.182 \\
3.5 & 0.185 & 0.185 & 0.199 & 0.183 \\
4 & 0.186 & 0.186 & 0.199 & 0.183 \\
\hline
Avg & 0.181 & 0.181 & 0.2 & 0.182 \\
\hline
\end{tabular}
\caption{Comparison of different portfolio optimization models in case of Simulated Data with 1000 samples (31 assets)}
\label{tab:1}
\end{table}

\begin{table}[h]
\centering
\captionsetup{justification=centering}
\begin{tabular}{||c|c|c|c|c||}
\hline
$\lambda$ & $SR_{Mark}$ & $SR_{Box}$ & $SR_{Ellip}$ & $SR_{Sep}$ \\
\hline
2 & 0.2 & 0.198 & 0.218 & 0.213 \\
2.5 & 0.203 & 0.204 & 0.218 & 0.213 \\
3 & 0.205 & 0.207 & 0.217 & 0.217 \\
3.5 & 0.208 & 0.209 & 0.216 & 0.222 \\
4 & 0.21 & 0.21 & 0.215 & 0.225 \\
\hline
Avg & 0.205 & 0.206 & 0.217 & 0.218 \\
\hline
\end{tabular}
\caption{Comparison of different portfolio optimization models in case of Simulated Data with same number of samples as market data (31 assets)}
\label{tab:2}
\end{table}

\begin{table}[h]
\centering
\captionsetup{justification=centering}
\begin{tabular}{||c|c|c|c|c||}
\hline
$\lambda$ & $SR_{Mark}$ & $SR_{Box}$ & $SR_{Ellip}$ & $SR_{Sep}$ \\
\hline
2 & 0.181 & 0.181 & 0.193 & 0.186 \\
2.5 & 0.181 & 0.181 & 0.192 & 0.193 \\
3 & 0.186 & 0.191 & 0.192 & 0.202 \\
3.5 & 0.194 & 0.195 & 0.191 & 0.209 \\
4 & 0.201 & 0.202 & 0.19 & 0.213 \\
\hline
Avg & 0.189 & 0.19 & 0.192 & 0.2 \\
\hline
\end{tabular}
\caption{Comparison of different portfolio optimization models in case of Market Data (31 assets)}
\label{tab:3}
\end{table}

\begin{table}[h]
\centering
\captionsetup{justification=centering}
\begin{tabular}{||c|c|c|c|c||}
\hline
$\lambda$ & $SR_{Mark}$ & $SR_{Box}$ & $SR_{Ellip}$ & $SR_{Sep}$ \\
\hline
2 & 0.185 & 0.185 & 0.245 & 0.191 \\
2.5 & 0.185 & 0.185 & 0.244 & 0.202 \\
3 & 0.19 & 0.192 & 0.244 & 0.213 \\
3.5 & 0.199 & 0.199 & 0.243 & 0.221 \\
4 & 0.207 & 0.207 & 0.242 & 0.226 \\
\hline
Avg & 0.193 & 0.194 & 0.244 & 0.21 \\
\hline
\end{tabular}
\caption{Comparison of different portfolio optimization models in case of Simulated Data with 1000 samples (98 assets)}
\label{tab:4}
\end{table}

\begin{table}[h]
\centering
\captionsetup{justification=centering}
\begin{tabular}{||c|c|c|c|c||}
\hline
$\lambda$ & $SR_{Mark}$ & $SR_{Box}$ & $SR_{Ellip}$ & $SR_{Sep}$ \\
\hline
2 & 0.192 & 0.192 & 0.235 & 0.216 \\
2.5 & 0.192 & 0.192 & 0.234 & 0.227 \\
3 & 0.202 & 0.203 & 0.233 & 0.233 \\
3.5 & 0.212 & 0.213 & 0.232 & 0.237 \\
4 & 0.221 & 0.222 & 0.232 & 0.239 \\
\hline
Avg & 0.204 & 0.205 & 0.233 & 0.23 \\
\hline
\end{tabular}
\caption{Comparison of different portfolio optimization models in case of Simulated Data with same number of samples as market data (98 assets)}
\label{tab:5}
\end{table}

\begin{table}[h]
\centering
\captionsetup{justification=centering}
\begin{tabular}{||c|c|c|c|c||}
\hline
$\lambda$ & $SR_{Mark}$ & $SR_{Box}$ & $SR_{Ellip}$ & $SR_{Sep}$ \\
\hline
2 & 0.175 & 0.173 & 0.195 & 0.193 \\
2.5 & 0.178 & 0.177 & 0.194 & 0.193 \\
3 & 0.18 & 0.181 & 0.194 & 0.193 \\
3.5 & 0.186 & 0.188 & 0.193 & 0.192 \\
4 & 0.191 & 0.192 & 0.193 & 0.192 \\
\hline
Avg & 0.182 & 0.182 & 0.194 & 0.192 \\
\hline
\end{tabular}
\caption{Comparison of different portfolio optimization models in case of Market Data (98 assets)}
\label{tab:6}
\end{table}

\begin{table}[h]
\centering
\captionsetup{justification=centering}
\begin{tabular}{||p{7cm}|p{3cm}|p{3cm}||}
\hline
& \#stocks = 31 & \#stocks = 98 \\
\hline
\#generated\textunderscore simulations = 1000  & 0.2    &0.244\\
\#generated\textunderscore simulations = $\zeta (<1000)$ & 0.218  & 0.233 \\
Market data & 0.2 & 0.194 \\
\hline
\end{tabular}
\caption{The maximum average Sharpe ratio compared by varying the number of stocks in different kinds of scenarios.}
\label{tab:no_stocks}
\end{table}

\begin{table}[h]
\centering
\captionsetup{justification=centering}
\begin{tabular}{||p{4cm}|p{4cm}|p{4cm}||}
\hline
& \#samples = 1000 & \#samples = $\zeta (< 1000)$ \\
\hline
\#stocks = 31  & 0.2    &0.218\\
\#stocks = 98 &   0.244  & 0.233 \\
\hline
\end{tabular}
\caption{The maximum average Sharpe ratio compared by varying the number of stocks in different kinds of scenarios.}
\label{tab:no_samples}
\end{table}

\begin{table}[h]
\centering
\captionsetup{justification=centering}
\begin{tabular}{||p{4cm}|p{4cm}|p{4cm}||}
\hline
& Simulated data & Real Market data \\
\hline
\#stocks = 31  & 0.218    &0.2\\
\#stocks = 98 &   0.244  & 0.194  \\
\hline
\end{tabular}
\caption{The maximum average Sharpe ratio compared by varying the type of the data in different kinds of scenarios.}
\label{tab:data_type}
\end{table}

\end{document}